\documentclass[aps,prl,twocolumn,amsmath,amssymb,showpacs,superscriptaddress,notitlepage,longbibliography]{revtex4-1}
\usepackage[colorlinks=true,linkcolor=blue,anchorcolor=red,citecolor=blue, urlcolor=blue]{hyperref}
\usepackage{bm}
\usepackage{graphicx}
\usepackage{color}

\begin{document}
	
	\title{Analytical solution for the surface states of antiferromagnetic topological insulator MnBi$_2$Te$_4$ }
	
	\author{Hai-Peng Sun}
	\affiliation{Department of Physics, Harbin Institute of Technology, Harbin 150001, China}
	\affiliation{Shenzhen Institute for Quantum Science and Engineering and Department of Physics, Southern University of Science and Technology (SUSTech), Shenzhen 518055, China}
	\affiliation{Shenzhen Key Laboratory of Quantum Science and Engineering, Shenzhen 518055, China}
	
	\author{C. M. Wang}
	\affiliation{Department of Physics, Shanghai Normal University, Shanghai, 200234, China}
	\affiliation{Shenzhen Institute for Quantum Science and Engineering and Department of Physics, Southern University of Science and Technology (SUSTech), Shenzhen 518055, China}
	\affiliation{Shenzhen Key Laboratory of Quantum Science and Engineering, Shenzhen 518055, China}
	
	\author{Song-Bo Zhang}
	\affiliation{Institute for Theoretical Physics and Astrophysics, University of Wurzberg, D-97074 Wurzburg, Germany}
	
	\author{Rui Chen}
	\affiliation{Shenzhen Institute for Quantum Science and Engineering and Department of Physics, Southern University of Science and Technology (SUSTech), Shenzhen 518055, China}
	\affiliation{Shenzhen Key Laboratory of Quantum Science and Engineering, Shenzhen 518055, China}

	\author{Yue Zhao}
	\affiliation{Shenzhen Institute for Quantum Science and Engineering and Department of Physics, Southern University of Science and Technology (SUSTech), Shenzhen 518055, China}
	
	\author{Chang Liu}
	\affiliation{Shenzhen Institute for Quantum Science and Engineering and Department of Physics, Southern University of Science and Technology (SUSTech), Shenzhen 518055, China}
	
	\author{Qihang Liu}
	\affiliation{Shenzhen Institute for Quantum Science and Engineering and Department of Physics, Southern University of Science and Technology (SUSTech), Shenzhen 518055, China}
	
	\author{Chaoyu Chen}
	\affiliation{Shenzhen Institute for Quantum Science and Engineering and Department of Physics, Southern University of Science and Technology (SUSTech), Shenzhen 518055, China}
	
	\author{Hai-Zhou Lu}
	\email{Corresponding author: luhz@sustech.edu.cn}
	\affiliation{Shenzhen Institute for Quantum Science and Engineering and Department of Physics, Southern University of Science and Technology (SUSTech), Shenzhen 518055, China}
	\affiliation{Shenzhen Key Laboratory of Quantum Science and Engineering, Shenzhen 518055, China}
	
	\author{X. C. Xie}
	\affiliation{International Center for Quantum Materials, School of Physics, Peking University, Beijing 100871, China}
	\affiliation{CAS Center for Excellence in Topological Quantum Computation, University of Chinese Academy of Sciences, Beijing 100190, China}
	\affiliation{Beijing Academy of Quantum Information Sciences, West Building 3, No. 10, Xibeiwang East Road, Haidian District, Beijing 100193, China}
	
	\date{\today }
	
	\begin{abstract}
		Recently, the intrinsic magnetic topological insulator MnBi$_2$Te$_4$ has attracted great attention. It has an out-of-plane antiferromagnetic order, which is believed to open a sizable energy gap in the surface states. This gap, however, was not always observable in the latest angle-resolved photoemission spectroscopy (ARPES) experiments. To address this issue, we analytically derive an effective model for the two-dimensional (2D) surface states by starting from a three-dimensional (3D) Hamiltonian for bulk MnBi$_2$Te$_4$ and taking into account the spatial profile of the bulk magnetization. Our calculations suggest that the diminished surface gap may be caused by a much smaller and more localized intralayer ferromagnetic order. In addition, we calculate the spatial distribution and penetration depth of the surface states, which indicates that the surface states are mainly embedded in the first two septuple layers from the terminating surface. From our analytical results, the influence of the bulk parameters on the surface states can be found explicitly. Furthermore, we derive a $\bf{k}\cdot \bf{p}$ model for MnBi$_2$Te$_4$ thin films and show the oscillation of the Chern number between odd and even septuple layers. Our results will be helpful for the ongoing explorations of the MnBi$_x$Te$_y$ family.
	\end{abstract}

	\maketitle
	
	{\color{blue}\emph{Introduction.}}-- A topological insulator has topologically protected conducting surface states and insulating interior \cite{Moore10nat, Hasan10rmp,Qi11rmp, Shen12book}. When it is doped with magnetic impurities \cite{Liu08prl,Yu10sci}, time reversal symmetry breaking can open a band gap in its surface states. The band gap is essential for the realization of the quantized anomalous Hall effect \cite{Yu10sci,Chang13sci}, which hosts dissipationless chiral edge states with potential applications in future devices.  However, as magnetic impurities inevitably introduce inhomogeneity and disorder, the temperature at which the quantized anomalous Hall effect can be observed is typically low ($<2$ K) \cite{Mogi15apl,Tokura19nrp}. The working temperature may be increased by using the magnetic extension method \cite{Otrokov17JETP, Otrokov172Dmat} or intrinsic magnetic topological insulators \cite{Mong10prb,Chowdhury19njpcm,Mong19nat,Rienks19nat}, where magnetism is part of the crystal. Recently, an intrinsic antiferromagnetic (AFM) topological insulator MnBi$_2$Te$_4$ was predicted \cite{Otrokov19nat, Zhang19prl, Li19sa} and has been verified by experiments \cite{Otrokov19nat, Gong19cpl, Chen19nc, Vidal19prbrc, Li19prx, Hao19prx, Chen19prx, Swatek20prb}.
Quantum anomalous Hall effect \cite{Deng20sci}, axion insulator to Chern insulator transition \cite{Liu20nmat}, and high Chern number quantum Hall effect without Landau levels \cite{Ge20nsr} have been reported in the MnBi$_2$Te$_4$ thin films.
Nevertheless, the latest experiments imply that the topological surface states in a bulk crystal of MnBi$_2$Te$_4$ may be gapless \cite{Hao19prx, Chen19prx, Swatek20prb}, in contrast to the earlier calculations and observations \cite{Otrokov19prl, Zhang19prl, Li19sa, Li19prbrc,  Wu19sa, Zeugner19chemmater, Vidal19prbrc, Otrokov19nat, Lee19prr, Gong19cpl, Chen19nc, Li19prx} (see Table.~\ref{Tab:GapSummary}).

\begin{figure}[!htpb]
	\centering
	\includegraphics[width=0.36\textwidth]{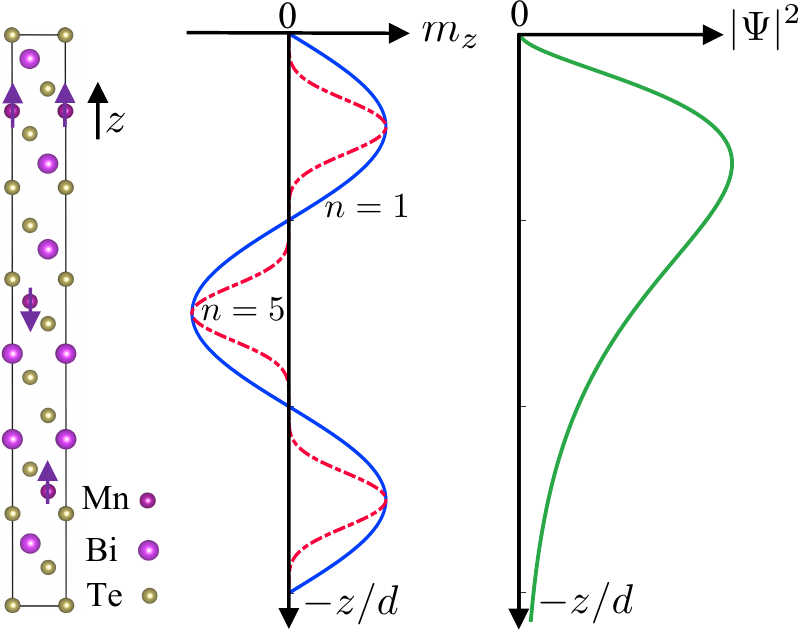}\\
	\caption{(Left) Septuple layers of MnBi$_2$Te$_4$ \cite{Villars16springer}. The purple arrows denote the direction of the ferromagnetic order in the Mn atomic layer. (Middle) The assumed profile of the bulk magnetization $m_z\sim \sin^n(\pi z/d)$. $n$ is an odd integer that controls the acting range of the intralayer ferromagnetic order and is the only tuning parameter in the calculation. $d$ is the thickness of a septuple layer. (Right) The distribution of the envelope function $\Psi$ of the surface states.}\label{Fig:lattice_mz_wavefunction}
\end{figure}

To understand why there is a wide range of the surface gap in the measurements, we derive analytically the effective model for the surface states as the analytical model can address the surface of a bulk crystal in terms of explicit expressions. Particularly, in the derivation we take into account the spatial profile of the bulk magnetization (Fig.~\ref{Fig:lattice_mz_wavefunction}). We show how the surface gap depends on the amplitude and acting range of the intralayer ferromagnetic order. We also summarize the estimated amplitude of the intralayer ferromagnetic order $m_0$ from the surface gap observed in the experiments (see Table~\ref{Tab:GapSummary}). In addition, we calculate the spatial distribution and penetration depth of the surface states, which indicates that the surface states are mainly embedded in the first two septuple layers from the terminating surface. Although we use the envelope function within the $\bf{k}\cdot \bf{p}$ approach, the order of magnitude of the penetration depth is consistent with the results obtained by the \emph{ab initio} calculations \cite{Shikin20sr}. We also derive a $\bf{k}\cdot \bf{p}$ model for MnBi$_2$Te$_4$ thin films. Our results will be helpful for the ongoing explorations of the MnBi$_x$Te$_y$ family.

\begin{table}[htb]
	\centering
	\caption{The gaps of the surface states of the AFM  MnBi$_2$Te$_4$ based on density functional theory (DFT) and angle-resolved photoemission spectroscopy (ARPES) and  the estimated amplitude of the intralayer ferromagnetic order $m_0$ from the experiments, where Eq.~(\ref{Eq:m_semi_infinite}) has been used and $n=1$ has been taken as an example. 
	}\label{Tab:GapSummary}
	\begin{ruledtabular}
		\begin{tabular}{lccc}
			Reference & Method & Surface gap (meV)  &  $m_0$ (meV) \\	
			Otrokov \emph{et al.} \cite{Otrokov19nat}& DFT & 88 & \\
			Otrokov \emph{et al.} \cite{Otrokov19nat}& ARPES & 70 & 218 \\				
			Lee \emph{et al.} \cite{Lee19prr} & ARPES & 85 & 265 \\	
			Zeugner \emph{et al.} \cite{Zeugner19chemmater} & ARPES & 100 & 311\\
			Chen \emph{et al.} \cite{Chen19nc}   & ARPES & $<25$ & $<78$ \\						
			Hao \emph{et al.} \cite{Hao19prx} & ARPES & $<3$ & $<9$  \\
			Li \emph{et al.} \cite{Li19prx} & ARPES & $<13.5$ & $<42$  \\
			Chen \emph{et al.} \cite{Chen19prx} & ARPES & $<2.5$  & $<8$ \\			
			Swatek \emph{et al.} \cite{Swatek20prb} & ARPES & $<2$ & $<6$ \\
			Shikin \emph{et al.} \cite{Shikin20sr} & ARPES & $15-67$ & $47-209$	
		\end{tabular}
	\end{ruledtabular}
\end{table}

{\color{blue}\emph{Effective model for surface states.}}-- The electronic structure of semi-infinite crystals can be calculated explicitly by the \emph{ab initio} calculations that include Green's functions. As an alternative way to calculate the surface states, the $\bf{k}\cdot \bf{p}$ approach can give qualitatively (even quantitatively) consistent results with the \emph{ab initio} calculations \cite{Shikin20sr}. A general approach for heterostructures containing the magnetically modulated doped topological insulator films can be found in Ref. \cite{Men'shov16JETP}, which provides a consistent interpretation of the well-known experimental findings \cite{Mogi15apl}. Starting from the three-dimensional (3D) Hamiltonian for the AFM1 (polarized along the $z$ direction) state of MnBi$_{2}$Te$_{4}$ \cite{Zhang19prl}, we derive the effective model for surface states in a semi-infinite geometry. The technique \cite{Shen12book} that we are using has been shown effective in the predictions of the quantum spin Hall effect in topological insulator thin films \cite{Lu10prb, Shan10njp,ZhangY10np} and 3D quantum Hall effect in topological semimetals \cite{WangCM17prl,ZhangC19nat}. The 3D Hamiltonian of AFM MnBi$_{2}$Te$_{4}$ reads

\begin{eqnarray} \label{Eq:MnBi2Te4}
\mathcal{H}(\mathbf{k}) &=& \epsilon_{0}(\mathbf{k}) +
\begin{bmatrix}
M(\mathbf{k}) & A_{1} k_{z}        & 0                 & A_{2} k_{-} \\
A_{1} k_{z}       & -M(\mathbf{k}) & A_{2} k_{-}       & 0 \\
0                 & A_{2} k_{+}        & M(\mathbf{k}) & -A_{1} k_{z} \\
A_{2} k_{+}       & 0                  & -A_{1} k_{z}      & -M(\mathbf{k})
\end{bmatrix}
\nonumber\\
&&+H_X,
\end{eqnarray}
where $k_{\pm}=k_{x} \pm i k_{y}$, $\epsilon_{0}(\boldsymbol{k})=  C_{0} + D_{1}k_{z}^{2} + D_{2}(k_{x}^{2}+k_{y}^{2})$ and $M(\boldsymbol{k})=M_{0} - B_{1}k_{z}^{2} - B_{2}(k_{x}^{2}+k_{y}^{2}) $. $C_0$, $D_i$, $M_0$, $B_i$ and $A_i$ are model parameters, where $i=1,2$. The basis of the Hamiltonian is \{$ |{P1_{z}^{+}, \uparrow}\rangle, |{P2_{z}^{-}, \uparrow}\rangle,|{P1_{z}^{+}, \downarrow}\rangle,|{P2_{z}^{-}, \downarrow}\rangle $\}.
Conventionally, the exchange field is treated as a constant $m_0$ in $H_X$. Unlike the previous works \cite{Lu10prb,Shan10njp}, we take into account the spatial profile of the exchange field in the AFM MnBi$_{2}$Te$_{4}$ for a realistic study. For simplicity, we consider the profile of the exchange field in terms of the sinusoidal function, and the exchange field $H_X$ reads
\begin{eqnarray}
H_X
= -m_0\ {\rm{sin}}^{n}(\frac{\pi}{d} z)\ \sigma_z \otimes \tau_0,
\end{eqnarray}
where $m_0$ is the amplitude of the intralayer ferromagnetic order, $d$ is the thickness of a septuple layer, $\sigma_{z}$ is the $z$ Pauli matrix for the spin degrees of freedom, $\tau_0$ is a 2$\times$2 unit matrix for the orbital degrees of freedom, and the magnetization energy along the $z$ direction $m_z = -m_0 {\rm{sin}}^{n}(\pi z/ d )$ is illustrated in Fig.~\ref{Fig:lattice_mz_wavefunction} for different $n$. The odd integer $n$ is used to model the localization of the intralayer ferromagnetic order. A larger $n$ indicates that the intralayer ferromagnetic order is more localized. Here, we focus on the out-of-plane magnetic moments according to the neutron diffraction results \cite{Yan19prm}.

We now consider a surface located at $z=0$ and derive the effective model for the surface states, as shown in Fig.~\ref{Fig:lattice_mz_wavefunction}. First, we find the solutions to the surface states at $k_x=k_y=0$ (the $\Gamma$ point), where the Hamiltonian becomes two $2\times 2$ blocks, denoted as $h(A_1)$
and $h(-A_1)$, respectively.
The trial solution to the eigen equation $h(A_1)\Psi=E\Psi$ can be written as $\Psi = (a,b)^T e^{\lambda z} $,
where $a$, $b$, and $\lambda$ are the coefficients and $E$ is the eigenenergy. There are two solutions to $\lambda$, denoted as $\lambda_1$ and $\lambda_2$, and the general solution is a linear combination of them, $\Psi(z)=\sum_{i=1}^2 (a_i,b_i)^T e^{\lambda_i z}$. For simplicity, we assume open boundary conditions on the surfaces. We note, however, the main features of surface states, such as their attenuation in the bulk, are generic and remain the same if we impose different boundary conditions. Thus, the envelope function $\Psi$ vanishes at $z=0$ and $z=-\infty$, which are applied to the general solution to obtain the eigenenergy at the $\Gamma$ point
$
E_0 = C_0 +M_0 D_1/B_1
$ and the envelope function
\begin{eqnarray}
\Psi^{\uparrow}(z)
\sim
\begin{bmatrix}
i A_{1}   \\
-D_{-} \lambda_{+}
\end{bmatrix}
( e^{ \lambda_{1} z}  - e^{ \lambda_{2} z} )
\end{eqnarray}
up to a normalization factor with $D_{-}=D_1-B_1$, and $\lambda_{+}=\lambda_1+\lambda_2$.
Replacing $A_1$ with $-A_1$ can give the eigenenergy and the envelope function $\Psi^{\downarrow}(z)$ of the other block $h(-A_1)$,  which is degenerate with $h(A_1)$ when there is no magnetization.

Next, we project the Hamiltonian at $\{k_x,k_y\}\ne 0$ into the basis expanded by $\Phi_1 =[\Psi^{\uparrow}(z), 0]^{T}$ and $\Phi_2 =[0, \Psi^{\downarrow}(z)]^{T}$, and obtain the 2D effective Hamiltonian for surface states localized at the surface,
\begin{eqnarray}
H_{\rm{surface}}
&=&
E_0 - Dk^2 +
\begin{bmatrix}
m & i \gamma k_{-} \\
-i \gamma k_{+} & -m
\end{bmatrix} ,\label{Eq:model}
\end{eqnarray}
where $D=\tilde{B_2}-D_2$, $\tilde{B_2}=B_2 \langle \Psi^{\uparrow}(z)|\sigma_z|\Psi^{\uparrow}(z)\rangle$, and $\gamma=-i A_2 \langle \Psi^{\uparrow}(z)|\sigma_x|\Psi^{\downarrow}(z)\rangle$. Note that this effective model is valid only in the vicinity of the $\Gamma$ point. $m$ is the effective magnetic moment. It can be found explicitly as \cite{supp}
\begin{eqnarray} \label{Eq:m_semi_infinite}
m
&=&
m_0^{\prime}  n! \pi^{n} d
\Bigg\{
\frac{2}{ \prod_{j=1}^{\frac{n+1}{2}} \left\{ d^2 \lambda_{+}^2 + [ (2j-1) \pi]^2 \right\}  } \nonumber \\
& & -\sum_{i=1}^{2} \frac{1}{\prod_{j=1}^{\frac{n+1}{2}} \left\{ 4 d^2 \lambda_{i}^2 + [ (2j-1) \pi]^2 \right\} }
\Bigg\}
\end{eqnarray}
for an arbitrary odd integer $n$, where the parameter $m_0^{\prime}=-m_0(A_{1}^2   + D_{-}^2 |\lambda_{+}|^2)|C|^2$ and $C$ is the normalization factor.

\begin{figure}[htpb]
	\centering
	\includegraphics[width=0.38\textwidth]{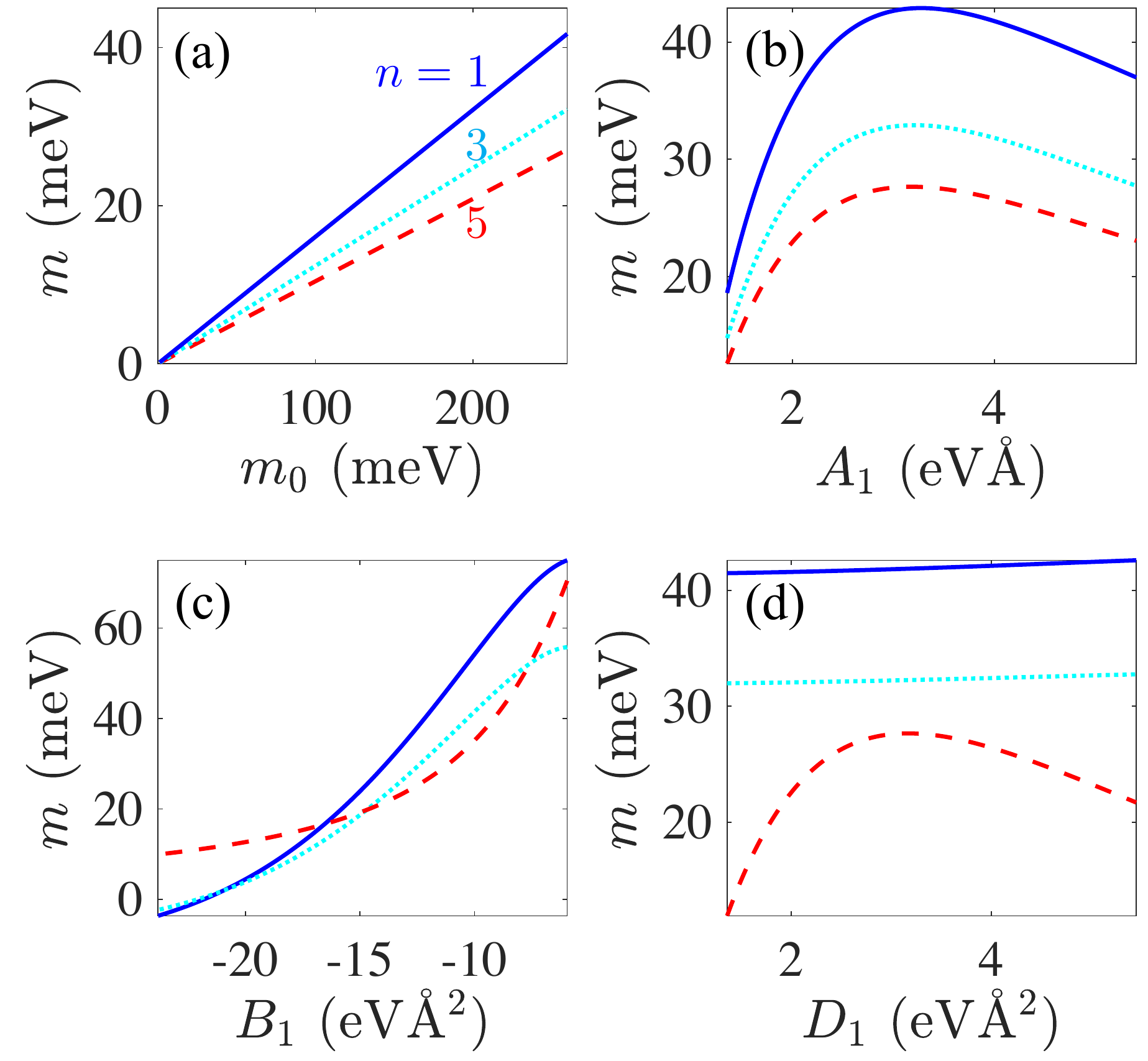}\\
	\centering \includegraphics[width=0.38\textwidth]{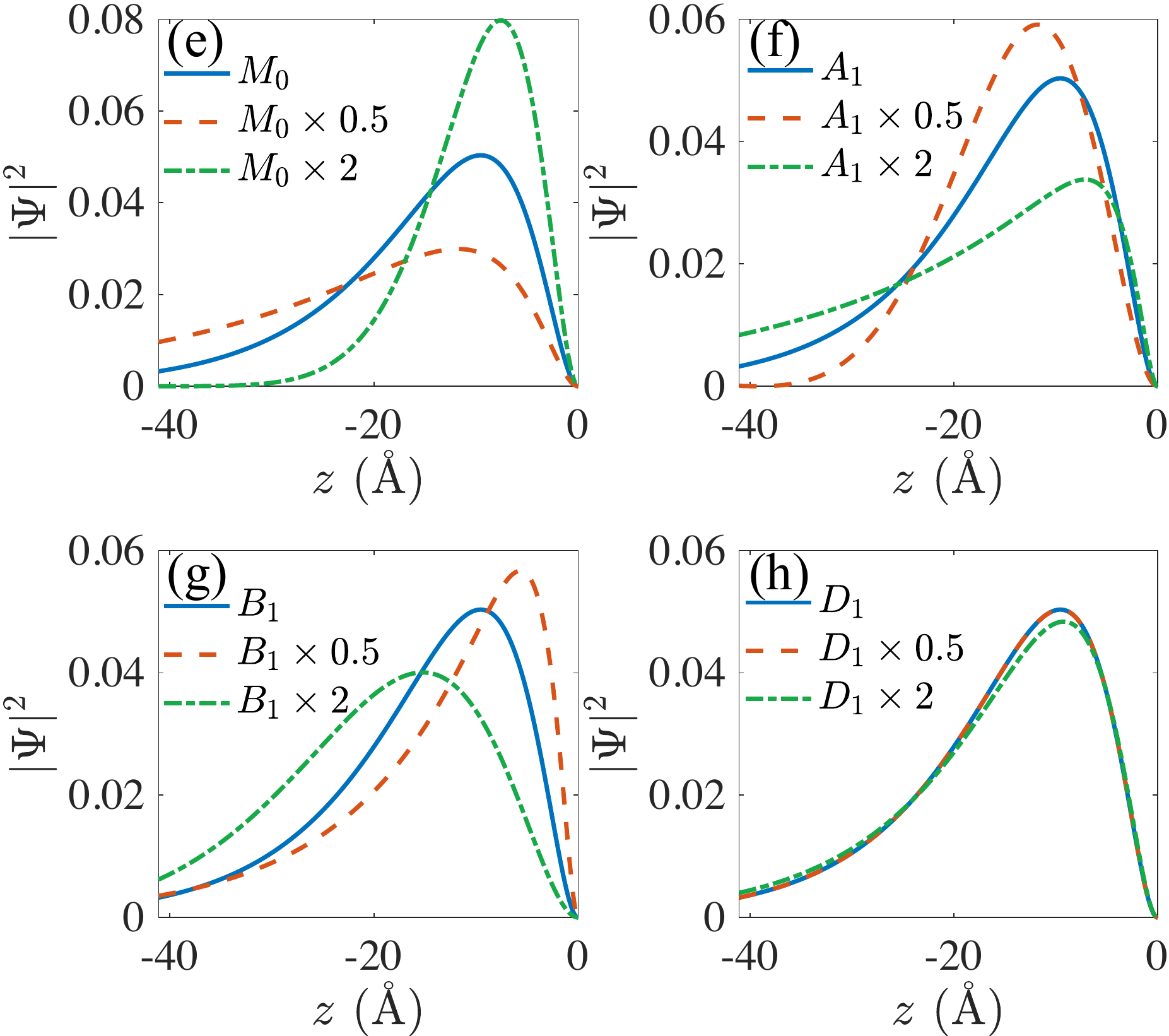}\\
	\caption{(a) The relation between the effective magnetic moment $m$ and the amplitude of the intralayer ferromagnetic order $m_0$ for different $m_z$ profiles in Fig.~\ref{Fig:lattice_mz_wavefunction}. [(b)--(d)] The influence of the variation of the parameters $ A_1, B_1$, and $D_1$ on the effective magnetic moment $m$, respectively. The blue solid, cyan dotted, and red dashed lines represent $n=1$,$3$, and 5, respectively. [(e)--(h)] The influence of the variation of the parameters $ M_0, A_1, B_1$, and $D_1$ on the distribution of the surface states, respectively. The parameters are adopted from \emph{ab initio} calculations \cite{Zhang19prl}, $C_{0}=-0.0048$ eV, $D_1 = 2.7232$ eV\AA$^2$, $D_2 = 17.0000$ eV\AA$^2$, $A_1 = 2.7023$ eV\AA, $A_2 = 3.1964$ eV\AA, $M_{0} = -0.1165$ eV, $B_1 = -11.9048$ eV\AA$^2$, $B_2 = -9.4048$ eV\AA$^2$, and $m_0 = 0.26$ eV.  }\label{Fig:m_parameterts}
\end{figure}

{\color{blue}\emph{Gap and penetration depth of surface states.}}--The two energy bands of the effective model, Eq.\ \eqref{Eq:model}, are given by
$
E_{\rm{surface}}^{\pm}
=
E_0 - Dk^2 \pm \sqrt{m^2 + \gamma^2 k^2}
$,
which shows that the surface states open a gap of magnitude 2$|m|$ at the $\Gamma$ point.
Using Eq. (5), we calculate the effective magnetic moment $m$ for increasing odd integers $n\in\{1,3,5,...\}$ and display $m$ as functions of different model parameters in Figs.~\ref{Fig:m_parameterts}(a)--\ref{Fig:m_parameterts}(d).
Figure~\ref{Fig:m_parameterts}(a) shows that the effective magnetic moment $m$ is reduced when the intralayer ferromagnetic order $m_0$ becomes smaller or more localized (for larger $n$).
In this sense, our calculations suggest that the diminished surface gap may be caused by a much smaller and more localized intralayer ferromagnetic order. It is worth mentioning that Shikin and collaborators proposed that the structural changes of the surface may cause the surface states moving inward, which will also result in a reduced effective magnetic moment \cite{Shikin20sr, Eremeev12njp}.

Besides, we calculate the spatial distribution $|\Psi(z)|^2$ and hence the penetration depth of surface states $\xi=|\langle\Psi|z|\Psi\rangle|$, using the parameters obtained by \emph{ab initio} calculations \cite{Zhang19prl}. As illustrated in Fig.~\ref{Fig:m_parameterts}(e)--\ref{Fig:m_parameterts}(h), the surface states are well localized at the surface, which have already been probed by the experiments \cite{Yuan20nanolett,Liang20prbrc}.
Importantly, we find that the penetration depth $\xi$ is about $16.2$ \AA, which is larger than the thickness of a septuple layer (about 13.7 \AA). This indicates that the surface states have already extended to the second septuple layer. Note that the penetration depth we estimated by using the envelope function is consistent with the results obtained by \emph{ab initio} calculations \cite{Shikin20sr,Zhang10njp}.

{\color{blue}\emph{Influences of the bulk on the surface gap.}}-- The parameters of the bulk Hamiltonian in Eq. (\ref{Eq:MnBi2Te4}), which are extracted from the \emph{ab initio} calculations \cite{Zhang19prl}, can change with external conditions, such as mechanical exfoliation, defects, or doping. Therefore,  we study the influence of the variation of the bulk parameters on the effective magnetic moment $m$ and surface gap $2|m|$. As $A_1$ increases, $m$ first increases and then decreases for different $n$, as shown in Fig.~\ref{Fig:m_parameterts}(b). The decrease of $m$ is due to the cancellation of the opposite magnetization between odd and even septuple layers, as the surface envelope function becomes more extended with increasing $A_1$, as shown in Fig.~\ref{Fig:m_parameterts}(f). 
By contrast, with increasing $B_1$, $m$ increases first slowly and then abruptly for $n=5$ and there are transitions from negative to positive $m$ for $n=1$ and 3, as illustrated in Fig.~\ref{Fig:m_parameterts}(c). With increasing $D_1$, $m$ increases slowly for $n=1$ and 3. Nevertheless, for $n=5$, $m$ first increases and then decreases, as illustrated in Fig.~\ref{Fig:m_parameterts}(d). The change of $m$ can also be understood from the distribution of the envelope function. As $B_1$ increases, the surface states move towards the surface plane, as depicted in Fig.~\ref{Fig:m_parameterts}(g). By contrast, as  $M_0$ or $A_1$ increases, the surface states move away from the surface plane, as depicted in Figs.~\ref{Fig:m_parameterts}(e) and \ref{Fig:m_parameterts}(f). In addition, as $D_1$ increases, the distribution of the surface states only changes slightly, as shown in Fig.~\ref{Fig:m_parameterts}(h).

\begin{figure}[htpb]
	\centering \includegraphics[width=0.38\textwidth]{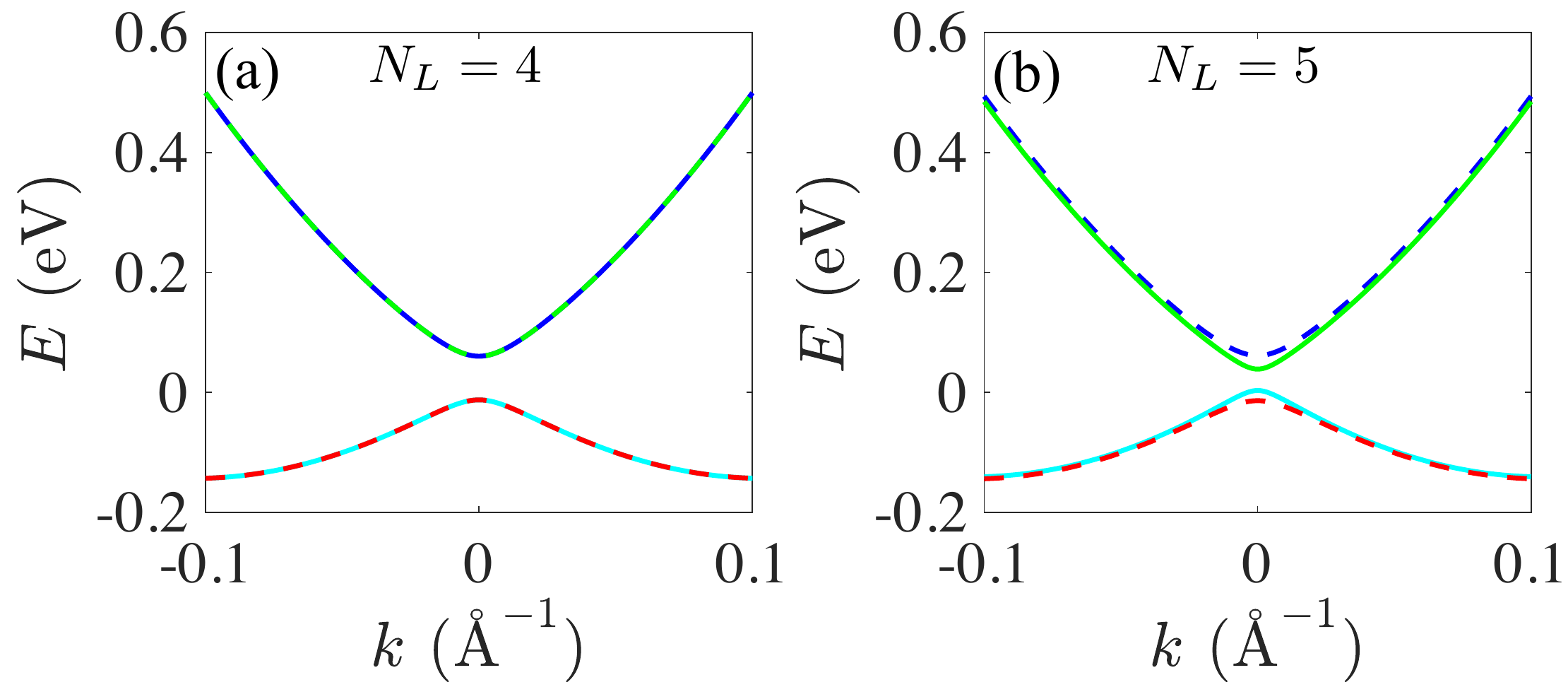}\\
	\centering \includegraphics[width=0.38\textwidth]{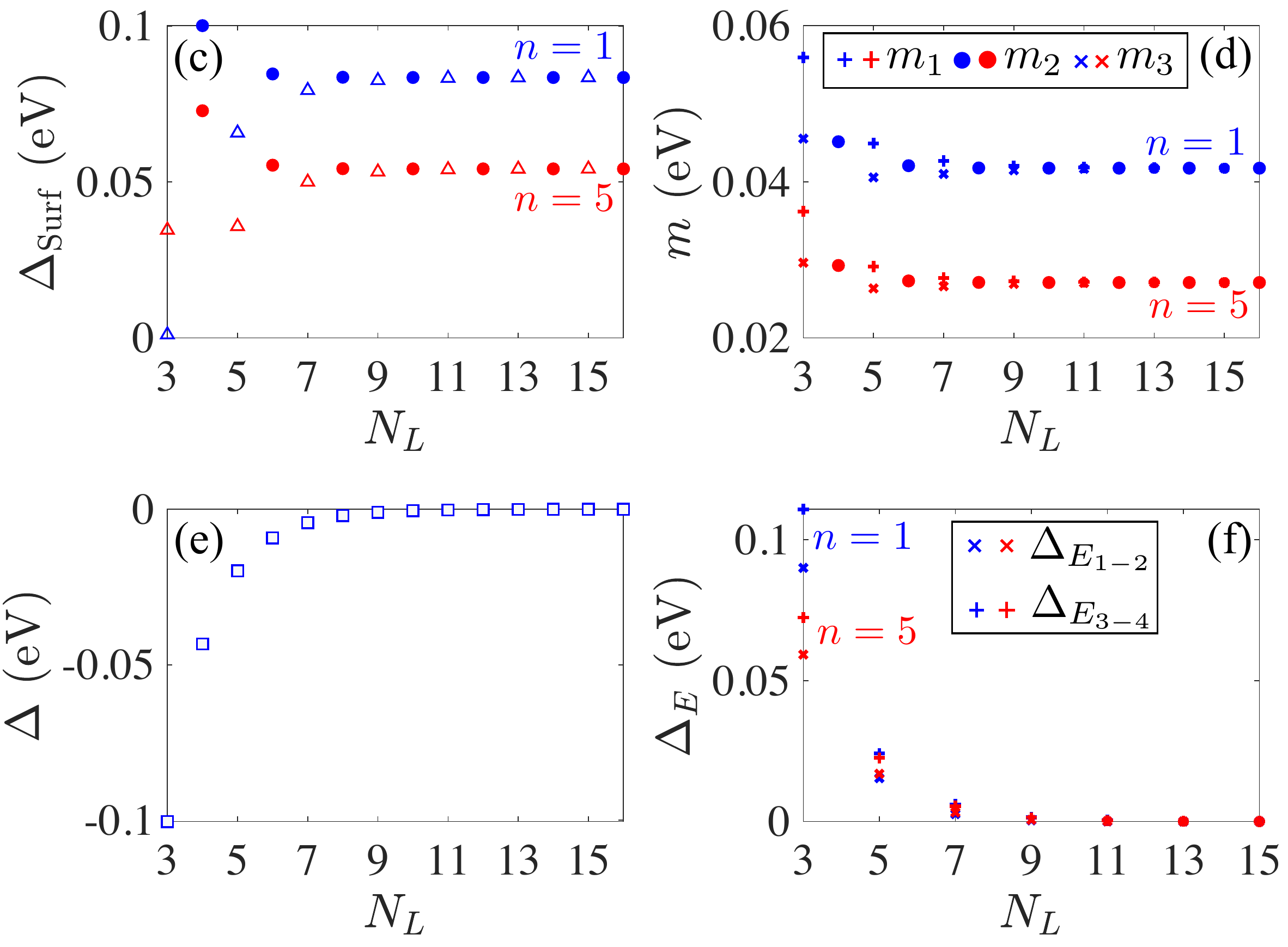}\\	
	\centering \includegraphics[width=0.38\textwidth]{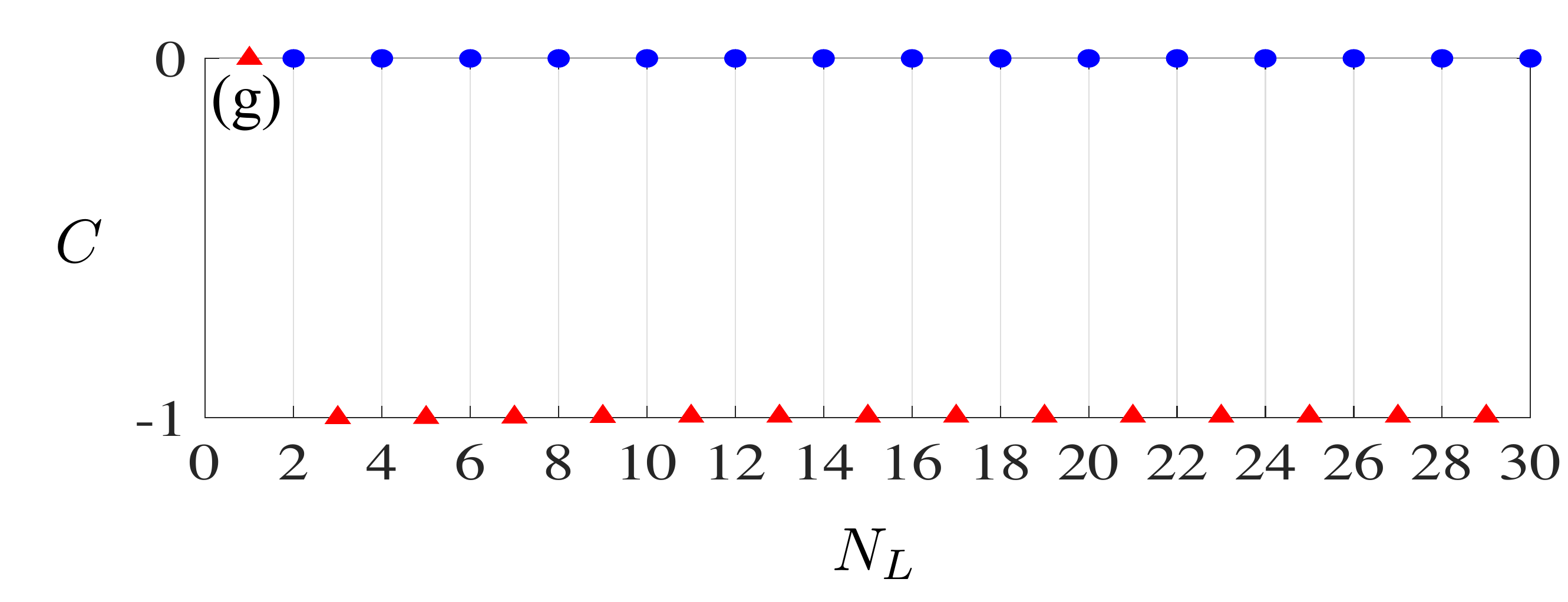}\\	
	\caption{The energy dispersion of even and odd septuple layers for (a) $N_L=4$ and (b) $N_L=5$. (c) The relation between the surface gap $\Delta_{\rm{Surf}}$ and the number of septuple layers $N_L$ for different $m_z$ profiles in Fig.~\ref{Fig:lattice_mz_wavefunction}. The blue solid circles and empty triangles represent $n=1$ and red solid circles and empty triangles represent $n=5$, respectively. (d) The effective magnetic moments $m_{1,2,3}$ as functions of $N_L$ for different $m_z$ profiles in Fig.~\ref{Fig:lattice_mz_wavefunction}. The blue plus signs, crosses, and solid circles represent $n=1$ and red plus signs, crosses, and solid circles represent $n=5$, respectively. (e) The finite-size gap $\Delta$ versus $N_L$.  (f) The energy difference of the upper two bands $\Delta_{E_{1-2}}$ and the energy difference of the lower two bands $\Delta_{E_{3-4}}$ versus $N_L$. (g) The Chern number $C$ versus $N_L$. For even $N_L$, the Chern number is zero. For odd $N_L \ge 3$, the Chern number is --1. The parameters are the same as in Fig.~\ref{Fig:m_parameterts}.  }\label{Fig:thin_films}
\end{figure}

{\color{blue}\emph{Effective model for MnBi$_2$Te$_4$ thin films.}}-- Above, we have discussed the gap at a single open surface in a semi-infinite geometry. Now, we turn to discuss the surface gap in a thin film with both top and bottom open surfaces. First, we derive the $\bf{k}\cdot \bf{p}$ model for MnBi$_2$Te$_4$ thin films. Following the same procedure as described above, we obtain the effective Hamiltonian for MnBi$_2$Te$_4$ thin films, which reads \cite{supp}
\begin{eqnarray}
&&H_{\rm{film}}
=
E_0 - D k^2 + \nonumber \\
&&
\begin{bmatrix}
h(k)+m_1 & i \gamma k_{-} & m_2 & 0 \\
-i \gamma k_{+} & -h(k)-m_3 & 0 & -m_2 \\
m_2 & 0 & -h(k)+m_3 & i \gamma k_{-} \\
0 & -m_2 & -i \gamma k_{+} & h(k)-m_1
\end{bmatrix} % \nonumber
\end{eqnarray}
where $E_0=(E_+ + E_-)/2$, $E_+ $, and $E_-$ are the eigenenergies at the $\Gamma$ point, $h(k)=\Delta/2 - b k^2$ with $k^2=k_x^2 + k_y^2$,  $\Delta = E_+ - E_-$ is the finite-size gap, and $m_{i}$ is the effective magnetic moment with $i=1, 2, 3$.
The eigenenergies of the surface states for even septuple layers are found as $E_{\rm{even}}=E_0 - D k^2 \pm \sqrt{(\Delta/2-bk^{2})^2 + \gamma^2 k^2 + m_2^2}$, which is two-fold degenerate due to $P_2 \Theta$ symmetry \cite{Li19sa}, as shown in Fig.~\ref{Fig:thin_films}(a). At the $\Gamma$ point, the surface gap for even septuple layers is $2\sqrt{(\Delta/2)^2 +m_2^2}$. It decreases as the thickness grows because both the finite-size gap $\Delta$ [Fig.~\ref{Fig:thin_films}(e)] and the effective magnetic moment $m_2$ [Fig.~\ref{Fig:thin_films}(d)] decrease with growing thickness. The surface gap and the effective magnetic moment $m_2$ finally saturate for larger $N_L$, as shown in Figs.~\ref{Fig:thin_films}(c) and 3(d).
The eigenenergies of the surface states for odd septuple layers are found as $E_{\rm{odd}}=E_0 - D k^2 + s(m_1-m_3)/2 \pm \sqrt{[ \Delta/2-bk^{2} + s(m_1+m_3)/2 ]^2 + \gamma^2 k^2}$ with $s=\pm 1$, which has no degeneracy due to the breaking of $\Theta$ and $P_2 \Theta$ symmetry \cite{Li19sa}, as shown in Fig.~\ref{Fig:thin_films}(b). At the $\Gamma$ point, the surface gap for odd septuple layers is $2\sqrt{[ \Delta/2 + (m_1+m_3)/2 ]^2}$.
For $n=1$ and $n=5$, the surface gap increases with growing $N_L$, and finally saturates for larger $N_L$. For the effective magnetic moment $m_3$, it first decreases, then increases, and finally saturates for larger $N_L$, as shown in Figs.~\ref{Fig:thin_films}(c)--(d).
Besides, the energy difference at the $\Gamma$ point between the upper two bands $\Delta_{E_{1-2}}$ is different from that of the lower two bands  $\Delta_{E_{3-4}}$ for odd septuple layers when $N_L<11$, due to the difference between $m_1$ and $m_3$, as shown in Figs.~\ref{Fig:thin_films}(d) and \ref{Fig:thin_films}(f). The Chern number is zero for all the even septuple layers and $N_L=1$ odd septuple layer, and is $-1$ for odd septuple layers with $N_L \ge 3$, as shown in Fig.~\ref{Fig:thin_films}(g). The oscillation of the Chern number in Fig.~\ref{Fig:thin_films}(g) is consistent with the nature of intrinsic antiferromagnetic topological insulators \cite{Zhang19prl,Otrokov19prl, Li19sa}. Note that $m_0=0.26$ eV has been taken in the calculation to be consistent with the results obtained by the \emph{ab initio} calculations \cite{Zhang19prl, Otrokov19prl, Li19sa}.

{\color{blue}\emph{Discussion.}}-- There are several experimental evidences that may support the idea that the intralayer ferromagnetic order becomes much smaller and more localized in real materials. The diminishment of the intralayer ferromagnetic order has been justified in recent experiments as the Mn atomic magnetic moment is about 1.14 $\mu_{B}$ in MnBi$_2$Te$_4$ thin films \cite{Gong19cpl} and around 3.8 $\mu_{B}$ in bulk MnBi$_2$Te$_4$ \cite{Hao19prx}, which is much smaller than the theoretically expected values of 4.6 $\mu_{B}$ \cite{Otrokov19nat,Otrokov19prl} or 5 $\mu_{B}$ \cite{Li19sa,Gong19cpl}.
Moreover, the resonance photoemission spectroscopy measurements have revealed that the density of Mn 3$d$ states is negligible in the energy range within 0.6 eV below the Fermi energy \cite{Li19prx,Vidal19prbrc}, which is similar to that in Mn-doped Sb$_2$Te$_3$ \cite{Islam18prb}. This implies a local nature of the Mn 3$d$ states in MnBi$_2$Te$_4$ \cite{Li19prx,Vidal19prbrc}.

\begin{acknowledgments}
	This work was supported by the National Natural Science Foundation of China (Grant No. 11534001, No. 11974249, and No. 11925402), the Strategic Priority Research Program of Chinese Academy of Sciences (Grant No. XDB28000000), Guangdong province (Grant No. 2016ZT06D348, and No. 2020KCXTD001), the National Key R \& D Program (Grant No. 2016YFA0301700), the Natural Science Foundation of Shanghai (Grant No. 19ZR1437300), Shenzhen High-level Special Fund (Grant No. G02206304, and No. G02206404), and the Science, Technology and Innovation Commission of Shenzhen Municipality (No. ZDSYS20170303165926217, No. JCYJ20170412152620376, and No. KYTDPT20181011104202253). The numerical calculations were supported by Center for Computational Science and Engineering of Southern University of Science and Technology.
\end{acknowledgments}

\end{document}